\documentclass[showpacs,prl,onecolumn,aps,superscriptaddress,preprintnumbers,nofootinbib]{revtex4}
\usepackage[T1]{fontenc}
\usepackage[latin9]{inputenc}
\usepackage{amsmath,amssymb}
\usepackage{epsfig}
\usepackage{bm} 
\usepackage{graphicx}
\usepackage{mathrsfs}
\usepackage{amsmath}
\usepackage{amsfonts}
\usepackage{epstopdf}
\usepackage{color}
\usepackage{pifont}
\usepackage{relsize}
\usepackage{mathtools}
\def\slashchar#1{\setbox0=\hbox{$#1$}     		
   \dimen0=\wd0                                 	
   \setbox1=\hbox{/} \dimen1=\wd1               	
   \ifdim\dimen0>\dimen1                        	
      \rlap{\hbox to \dimen0{\hfil/\hfil}}      	
      #1                                        	
   \else                                        	
      \rlap{\hbox to \dimen1{\hfil$#1$\hfil}}   	
      /                                         	
   \fi}

\renewcommand{\vec}{\boldsymbol}
\newcommand{\beq}{\begin{equation}}
\newcommand{\eeq}{\end{equation}}
\newcommand{\bea}{\begin{eqnarray}}
\newcommand{\eea}{\end{eqnarray}}
\newcommand{\ba}{\begin{array}}
\newcommand{\ea}{\end{array}}

\def\eq#1{{Eq.~(\ref{#1})}}
\def\fig#1{{Fig.~\ref{#1}}}
\newcommand{\bas}{\bar{\alpha}_S}
\newcommand{\as}{\alpha_S}
\newcommand{\nn}{\nonumber}

\newcommand{\Lb}{\left(}
\newcommand{\Rb}{\right)}
\newcommand{\h}{\frac{1}{2}}
\newcommand{\rv}{\vec{r}}

\newcommand{\bv}{\vec{b}}
\newcommand{\pom}{I\!\!P}

\newcommand{\intl}{\int\limits}
\begin{document}

\title{Scattering amplitude in QCD: summing large Pomeron loops }

\author{Eugene Levin}
\email{leving@tauex.tau.ac.il}
\affiliation{Department of Particle Physics, Tel Aviv University, Tel Aviv 69978, Israel}

\date{\today}


\begin{abstract}
In this paper we  show that the  sum of enhanced  BFKL Pomeron loop diagrams generates the scattering amplitude, which turns out to be much smaller,  than in the case of deep inelastic scattering. We use the simplified BFKL kernel in the leading twist approximation, which reproduces the main features of the scattering amplitude in the deep inelastic scattering(DIS).
For such kernel the results are highly unexpected and they contradict
 (i)  the solution to the Balitsky-Kovchegov(BK) equation for the scattering amplitude; (ii)  the idea 
  that the scattering amplitude stems from rare fluctuation and it has the same form as in DIS  ; and (iii) the numerical simulations.  
We sincerely hope, that we made a mistake, which we failed to note,   and which our reader will  find. If not , we need to reconsider our view on the sum of the BFKL Pomeron loops and accept that their summing will lead to large contribution of the rare    configurations in  CGC approach to the  scattering amplitude.

 \end{abstract}
\maketitle

\vspace{-0.5cm}
\tableofcontents

\section{My several words and results}
 In this paper we sum the large BFKL Pomeron loops for dipole-dipole scattering at high energies.
 The main equations for the scattering amplitude and  dipoles densities we have discussed in our previous paper \cite{LE1} . However, in this paper we use the different approximation to the BFKL\cite{BFKL}\footnote{BFKL stands for Balitsky, Fadin,Kuraev and Lipatov.} kernel: the BFKL kernel in the leading twist approximation\cite{LETU} (see also \cite{KOLEB,CLMS}). For deep inelastic scattering (DIS) processes this kernel generates the scattering amplitude which shows the geometric scaling behaviour \cite{GS}  and has the following form\cite{LETU}:
 \beq\label{I1}
 N^{\rm DIS} \Lb z= r^2 Q^2_s\Lb Y, b\Rb\Rb\,\,=\,\,1\,\,-\,\,C(z)\exp\Lb - \frac{z^2}{8}\Rb
 \eeq
 In \eq{I1}, $r$  is the size  and $Y$   is the rapidity    of the interacting dipole, and $b$ is the  impact parameter of scattering. In Ref.\cite{LE1} it is shown that the widely used  diffusion approximation for the BFKL kernel leads to the scattering amplitude which decreases at large $Y$ and, because of this,  it cannot be used for modeling both the DIS and the interaction of two dipoles at high energies. The plausible interpretation of \eq{I1} is given in Refs.\cite{IAMU,IAMU1}. It consists of two parts. First,   the typical contributions to the S- matrix at high energies are very small and can be neglected. Second,
  the
rare fluctuations   estimated in Ref.\cite{IAMU},  lead to the scattering amplitude of \eq{I1}-type, generating $\exp\Lb - C\,z^2\Rb$ suppression.

The main result of this paper is to show that the sum of large Pomeron loops leads to the scattering amplitude of two dipoles  which at high energy has the following form:
\beq\label{I2}
 N^{\rm  dipole-dipole} \Lb z= r^2 Q^2_s\Lb Y,R, b\Rb\Rb\,\,=\,\,1\,\,-\,\,C(z)\exp\Lb - \frac{z}{2}\Rb
 \eeq  
 where $R$ is the size of the target dipole.

 \eq{I2} means that  the fluctuations , which give the main contributions to S-matrix, turn out to be not small for scattering processes. This feature we have seen in the unitarity models in zero transverse dimensions\cite{KLLN}.
 This unexpected result  contradicts not only the arguments of Ref.\cite{IAMU} but also the experience with the DIS. Frankly speaking the main motivation for writing this paper is a hope that a reader who is smarter than me, will find a mistake in our estimates. It worthwhile mentioning that the arguments of Ref.\cite{IAMU} are partly based
 on the numerical simulations in Ref.\cite{MUSA}, which have to be taken with a grain of salt since they are based on the BFKL kernel in the diffusion approximation which  cannot be trusted as it has been shown in Ref.\cite{LE1}.
 
 On the other hand, this is the first attempt to sum Pomeron loops in a reasonable approximation for the BFKL kernel,  which could be instructive.  Summing Pomeron loops  has been one of the  difficult problems in the Color Glass Condensate (CGC) approach, without solving which we cannot consider the dilute-dilute and dense-dense parton densities collisions. As recently shown\cite{KLLL1,KLLL2}, even the Balitsky-Kovchegov (BK) equation, that governs the dilute-dense parton density  scattering (deep inelastic scattering (DIS) of electron with proton), has to be modified due to  contributions of  Pomeron loops.  However, in spite of intensive work 
\cite{BFKL,MUSA,LELU,LIP,KO1,LE11,RS,KLremark2,SHXI,KOLEV,nestor,LEPRI,LMM,LEM,MUT,IIML,LIREV,LIFT,GLR,GLR1,MUQI,MUDI,Salam,NAPE,BART,BKP,MV, KOLE,BRN,BRAUN,BK,KOLU,JIMWLK1,JIMWLK2,JIMWLK3, JIMWLK4,JIMWLK5,JIMWLK6,JIMWLK7,JIMWLK8,AKLL,KOLU11,KOLUD,BA05,SMITH,KLW,KLLL1,KLLL2,kl,LEPR}, this problem  has  not been solved.  Therefore, the result of this paper, if it is correct,  is a certain breakthrough. 
     \begin{figure}[ht]
    \centering
  \leavevmode
      \includegraphics[width=12cm]{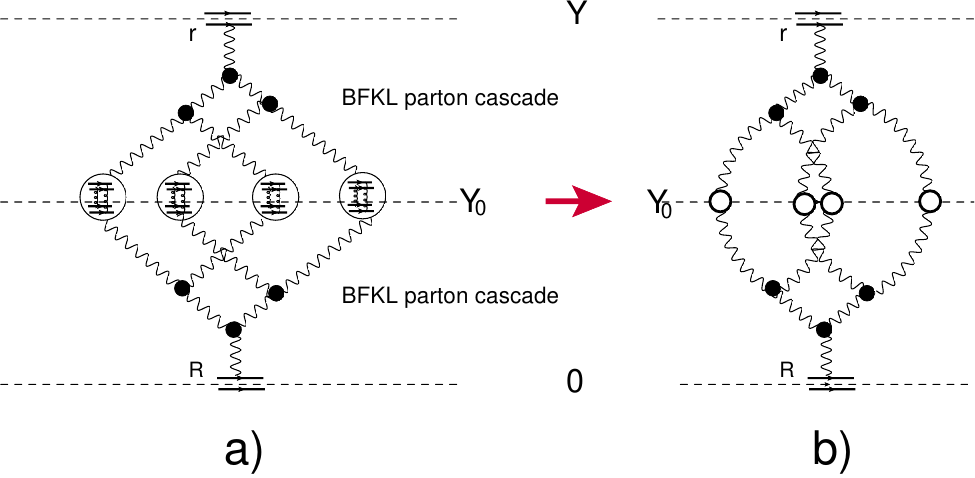}  
      \caption{ Summing  large Pomeron loops. The wavy lines denote the  BFKL Pomeron exchanges in \fig{mpsi}-a and  in \fig{mpsi}-b the Pomeron in the two dimensional Pomeron calculus with
      the Green's function $G_{\pom}(Y) = \exp\Lb \Delta\,Y\Rb$. The black circles stand for the triple Pomer
on vertices in both figures, which are equal to $\Delta$ In \fig{mpsi}-b, while the white circles denote the amplitude $\gamma$. In \fig{mpsi}-a we show the dipole-dipole scattering amplitude in the Born approximation of perturbative QCD in the circles.
 }
\label{mpsi}
   \end{figure}
 We sum the large Pomeron loops using the $t$-channel unitarity, which has been rewritten in the convenient form for the dipole approach to CGC in Refs.\cite{MUSA,Salam,IAMU,KOLEB,MUDI,LELU,KO1,LE11}.(see \fig{mpsi}-a).
 The analytic expression takes the form
       \cite{LELU,KO1,LE1}:  \bea \label{MPSI}
     && A\Lb Y, r,R ;  \vec{b}\Rb\,=\\
     &&\,\sum^\infty_{n=1}\,\Lb -1\Rb^{n+1}\,n!\int  \prod^n_i \frac{d^2 r_i}{r^4_i}\,\frac{d^2\,r'_i}{r'^4_i }\,d^2 b'_i 
     \int \!\!d^2 \delta b_i\, \gamma^{BA}\Lb r_1,r'_i, \vec{b}_i -  \vec{b'_i}\equiv \delta \vec{b} _i\Rb 
    \,\,\bar{\rho}_n\Lb Y - Y_0, \{ \vec{r}_i,\vec{b}_i\}\Rb\,\bar{\rho}_n\Lb Y_0, \{ \vec{r}'_i,\vec{b}'_i\}\Rb \nn
      \eea
  $\gamma^{BA}$ is the scattering amplitude of two dipoles in the Born approximation of perturbative QCD. 
   In \eq{MPSI} $\vec{b}_i\,\,=\,\,\vec{b} \,-\,\vec{b'}_i$. 
   
   In the next section we discuss the simplified BFKL kernel, that we use  in this paper, and solve the evolution equations for the dipole densities $\rho_n\Lb Y, r, \{r_1,b_i\}\Rb$. It turns out that   at high energies these densities  are
   \beq \label{I3}
   \rho_n\Lb Y, r, \{r_1,b_i\}\Rb \,\,\,=\,\,\,{\rm C_n} \prod_{i=1}^n \rho_1\Lb Y, r, r_i,b_i\Rb
   \eeq  
  where ${\rm C_n}$ is a constant. $\rho_1$ is the solution to the BFKL equation which we  consider in details.
  
  In section 3 we estimate the scattering amplitude using \eq{MPSI} and \eq{I3} and obtain \eq{I2}. In our presentation we try to clear all assumptions that we use simplifying  a search for possible mistakes.
  In conclusions we summarize our results and discuss possible flaw in our approach,

    \begin{boldmath}
    \section{Dipole densities $\rho_n\Lb r,b, \{r_i,b_i\}\Rb$ } 
    \end{boldmath}

    \subsection{ The BFKL leading twist kernel}
    
       \subsubsection{ The BFKL kernel: generalities}   
The BFKL evolution equation has the following for  the scattering amplitude  of the dipole with the size $r_1$ at rapidity $Y$  and impact parameter $b_1$ $N\Lb Y, r,  b\Rb$\cite{BFKL,MUDI}
\bea \label{BFKL1}
&&\frac{\partial \,N\Lb Y, \vec{r}_1,  \vec{b_1}\Rb}{ 
\bas\,\partial\,Y}\,\,=\\
&&\,\,\int\,\frac{d^2\,r'}{2\,\pi}\,
K\Lb \vec{r}',\vec{r}_1 - \vec{r'}|\vec{r}_1\Rb\,\Bigg\{N\Lb Y, r',  \vec{b_1} - \h(\vec{r_1} - \vec{r}')\Rb 
\,\,+\,\,N\Lb Y, \vec{r}_1 - \vec{r}',  \vec{b}_1 - \h \vec{r}'\Rb\,\,-\,\,N\Lb Y, \vec{r}_1,  \vec{b}_1\Rb\Bigg\}\nn
\eea
with 
\beq \label{K}
K\Lb \vec{r}',\vec{r}_1 - \vec{r'}|\vec{r}_1\Rb\,\,=\,\,\frac{r^2_1}{( \vec{r}_1 \,-\,\vec{r}')*2\,r'^2}
\eeq
   In Ref.\cite{LIP} it is found 
     that the eigenfunction of the BFKL equation has the following form

\beq \label{EIGENF}
\phi_\gamma\Lb \vec{r} , \vec{r}_1, \vec{b}_1\Rb\,\,\,=\,\,\,\Lb \frac{ r^2\,r_1^2}{\Lb \vec{b}_1  + \h(\vec{r} - \vec{r}_1)\Rb^2\,\Lb \vec{b}_1  -  \h(\vec{r} - \vec{r}_1)\Rb^2}\Rb^\gamma\,\,\xrightarrow{b_1\,\gg\,r,r_1}\,\,\Lb \frac{ r^2\,r_1^2}{b_1^4}\Rb^\gamma\,\,\equiv\,\,e^{\gamma\,\xi}\eeq
for any kernel which satisfies the conformal symmetry. In \eq{EIGENF} $r$ is the size of the initial dipole at $Y=0$  while $r_1$ is the size of the dipole with rapidity $Y$.  For the simplified kernel which we are going to consider, we will show
 the typical $b_i$ in $\bar{\rho}_n$ in \eq{BFKL1} is large. Hence we can use the variable $\xi$ from \eq{EIGENF}.

For the kernel of the LO BFKL equation (see \eq{K}) the eigenvalues  take the form:
\beq \label{CHI}
\omega\Lb \bas, \gamma\Rb\,\,=\,\,\bas\,\chi\Lb \gamma \Rb\,\,\,=\,\,\,\bas \Lb 2 \psi\Lb 1\Rb \,-\,\psi\Lb \gamma\Rb\,-\,\psi\Lb 1 - \gamma\Rb\Rb\eeq
where $\psi(z)$  is the Euler psi-function $\psi\Lb z\Rb = d \ln \Gamma(z)/d z$,    $\bas = N_c \as/\pi$, where $N_c$ is the number of colours. The general solution to \eq{BFKL1} takes the form:

\beq \label{SOL2}
N(Y; r_1, b)\,\,=\,\,\int\limits^{\epsilon + i \infty}_{\epsilon - i \infty} \frac{d\,\gamma}{2\,\pi\,i} 
\int\limits^{\epsilon + i \infty}_{\epsilon - i \infty} \frac{d\,\omega}{2\,\pi\,i}\frac{1}{\omega\,-\, \omega\Lb \bas, \gamma\Rb} e^{\omega\,Y\,\,+\,\,\gamma\,\xi} \phi_{in}\Lb \gamma\Rb\,\,=\,\,
  \,\int\limits^{\epsilon + i \infty}_{\epsilon - i \infty} \frac{d\,\gamma}{2\,\pi\,i} 
 e^{\omega\Lb \bas, \gamma\Rb\,Y\,\,+\,\,\gamma\,\xi} \phi_{in}\Lb \gamma\Rb   
   \eeq  
    Function $\phi_{in}\Lb \gamma\Rb$ has  been found from the initial conditions at $Y=0$\cite{LIP}(see also Refs.\cite{MUDI,LIREV}): $\phi_{in} \,=\,i \nu/\pi$, where $\gamma = \h + i \nu$.

     \begin{boldmath}
     \subsubsection{
      $\rho_1$ for BFKL leading twist kernel. }
      \end{boldmath}

      In this section, we  are going to discuss the solution for $\rho_1 \Lb r,r_1,b_1,Y\Rb $ for  the simplified BFKL kernel    suggested in     Ref.\cite{LETU} . This kernel describes the high energy asymptotic solution of the nonlinear BK equation and leads to the geometric scaling behaviour.: 
    \bea \label{SIMKER}
\chi\Lb \gamma\Rb\,\,=\,\, \left\{\begin{array}{l}\,\,\,\frac{1} {1\,-\,\gamma}\,\,\,\,\,\,\,\,\,\,\mbox{for}\,\,z = \xi + \lambda\,\bas\,Y \,>\,0,\,\,\,\,\,\,\mbox{summing} \Lb z\Rb^n;\\ \\
\,\,\,\frac{1}{\gamma}\,\,\,\,\,~~~~~~\mbox{for}\,\,\,z \,<\,0,\,\,\,\,\,\mbox{summing}
\Lb \xi\Rb^n;\\  \end{array}
\right.
\eea
 where $ \xi $ has been introduced in \eq{EIGENF} and $\lambda  = 4 $ for this kernel.  
 
 Since this kernel sums log contributions it corresponds to leading twist   term of the full BFKL kernel. It has a very simple form in the coordinate representation \cite{LETU}:
 
\beq \label{K2}
\int \, \displaystyle{ K\Lb \vec{r}',\vec{r} - \vec{r'}|\vec{r}\Rb}\,d^2 r' \,\rightarrow
\,\frac{\bas}{2}\, \int^{r^2}_{1/Q^2_s(Y,b)} \frac{ d r'^2}{r'^2}\,\,+\,\,
\frac{\bas}{2}\, \int^{r^2}_{1/Q^2_s(Y, b)} \frac{ d |\vec{r} - \vec{r}'|}{|\vec{r}  - \vec{r}'|^2}\,\,
 =\,\,\frac{\bas}{2}\, \int^{\xi}_{\xi_s} d \xi_{r'} \,\,+\,\,\frac{\bas}{2}\, \int^{\xi}_{\xi_s} d \xi_{\vec{r} - \vec{r}'}\eeq
 where $\xi_{r'} \,=\,\ln \Lb \frac{r'^2\,r^2_1}{b^4}\Rb$
 and $\xi_s\,=\,- \lambda\, \bas\,Y$ for the scattering of the  dipole $r'$ with the dipole $r_1$.
 Note, that  the  logarithms
   originate from the decay of a large size dipole, into one small
 size dipole  and one large size dipole\cite{LETU}.  However, the size of the
 small dipole is still larger than $1/Q^2_s  = \frac{b^4}{r^2_1}\,e^{-\lambda Y}  $.      The advantage of this kernel (\eq{SIMKER} ,\eq{K2}) is that it leads to the solution to the BK equation , which shows the geometric scaling behaviour (see Ref. \cite{GS,LETU})  and which leads to  saturated scattering amplitude that reaches the unitarity limit\cite{LETU}. For completeness of presentation we will discuss  these features in the next subsection.
 
 It turns out that  the equation for $\rho_1$ \cite{LE1} is the BFKL equation which for our simplified kernel takes the form: 
      \beq \label{RH1}
\frac{\partial \,\bar{\rho}_1(Y; \xi)}{ 
\partial\,Y}\,\,=\,\bas\,\int^{\xi}_{\xi_s} d \xi'\,
\bar{\rho}_1\Lb Y,\xi'\Rb;~~~~~~~\lambda \frac{d^2 \,\bar{\rho}_1(z)}{ 
d z^2}\,\,=\,\,\bar{\rho}_1(z) \eeq    
  with the obvious solution:
   \beq \label{RH11}    
\bar{\rho}_1(z)\,\,=\,\,{\rm C_1} e^{\sqrt{\frac{1}{\lambda}} \,z}  \,\,+\,\,\,{\rm C_2} e^{-\sqrt{\frac{1}{\lambda}}\,z } 
    \eeq
    Recall that  $\lambda =  4$  in \eq{RH1}.
    
    It should be emphasized that we find the solution which leads to geometric scaling behaviour.  For $\rho_1\Lb Y, z\Rb $  \eq{RH1} can be reduced to
  \beq \label{RH12}
\frac{\partial^2 \,\bar{\rho}_1(Y; z)}{ 
\partial\,Y\,\partial z}\,\,=\,\bas\,
\bar{\rho}_1\Lb Y,z \Rb  
\eeq
with the solution: $C_1 \,I_0\Lb 2\sqrt{\bas \,z\,Y}\Rb\,+\,C_2\,I_0\Lb- \,2\sqrt{\bas \,z\,Y}\Rb$.
  All results of this paper can be rewritten for this solution but we believe that the general arguments for the geometric scaling behaviour have been given in Refs.\cite{GS,LETU}  and it has been shown that this behaviour describes the HERA data on DIIS\cite{GS}. 

  It is instructive to note, that
   using the solution of \eq{RH11}  for $z \,<\,0$  we reproduce the 
  the general solution in the vicinity of the saturation scale\cite{MUT,IIML} $ \bar{\rho}_1(z)\,= \,N_0 \exp\Lb \bar{\gamma} z\Rb$ with $\bar{\gamma} = \h$.   For our kernel, we obtain
  \beq \label{RH12}    
    \bar{\rho}_1(z)\,\,=\,\,N_0 \,\,e^{ \h z}     \eeq      
    
     \begin{boldmath}
     \subsubsection{
    BK equation for BFKL leading twist kernel. }
      \end{boldmath}

The scattering amplitude of the colourless dipole with the size $x_{01}$ which determines the DIS cross section, satisfies the Balitsky-Kovchegov (BK) non-linear equation\cite{BK}:
 \beq \label{BK}
 \frac{\partial N_{01}}{\partial Y}\,=\,\bas\int \frac{d^2\,x_{02}}{2 \pi} \frac{ x^2_{01}}{x^2_{02}\,x^2_{12}}\Big\{ N_{02} + N_{12} - N_{02}N_{12} - N_{01}\Big\}
 \eeq
 where $N_{ik}=N\Lb Y, \vec{x}_{ik},\vec{b}\Rb$ is the scattering amplitude of the dipoles with size $x_{ik}$ and with rapidity $Y$ at the impact parameter $\vec{b}$. In \eq{BK} we assume that $b \gg r_{ik}$.
 
 For general BFKL kernel it has been proven several features of this equation. First,  it turns out that 
  \eq{BK} leads to a new dimensional scale: saturation momentum\cite{GLR}  which has the following $Y$ dependence\cite{GLR,MUT,MUPE}:
 \beq \label{QS}
 Q^2_s\Lb Y, b\Rb\,\,=\,\,Q^2_s\Lb Y=Y_0, b\Rb \,e^{\bas\,\kappa \,Y,-\,\,\frac{3}{2\,\gamma_{cr}} \ln Y }
 \eeq 
 where $Y_0$ is the initial value of rapidity and $\kappa$ and $\gamma_{cr}$   are determined by the following equations\footnote{$\chi\Lb \gamma\Rb$ is the BFKL kernel\cite{BFKL} in anomalous dimension ($\gamma$) representation.$\psi$ is the Euler psi -function (see Ref.\cite{RY} formula {\bf 8.36}). }:
  \beq \label{GACR}
\kappa \,\,\equiv\,\, \frac{\chi\Lb \gamma_{cr}\Rb}{1 - \gamma_{cr}}\,\,=\,\, - \frac{d \chi\Lb \gamma_{cr}\Rb}{d \gamma_{cr}}~
\eeq
where $\chi\Lb \gamma\Rb$ is defined in \eq{CHI}.

   Second, it is shown that in the vicinity  of the saturation scale the scattering amplitude take the following form\cite{MUT,MUPE}:  
  \beq \label{VQS}
 N_{01}\Lb  z\Rb\,\,\,=\,\,\,\mbox{Const} \Lb x^2_{10}\,Q^2_s\Lb Y\Rb\Rb^{\bar \gamma}
 \eeq
 with $\bar \gamma = 1 - \gamma_{cr}$. In \eq{VQS} we introduce a new variable $z$, which is equal to:
  \beq \label{z}
 z\,\,=\,\,\ln\Lb x^2_{01}\,Q^2_s\Lb Y, b\Rb\Rb\,\,\, =\,\,\,\,\bas\,\kappa \,\Lb Y\,-\,Y_A\Rb\,\,+\,\,\xi
 \eeq  
 where $\xi = \ln x^2_{01}$. Note that  we neglected the term $\frac{3}{2\,\gamma_{cr}} \ln Y$ in \eq{QS}. 
 
 Third, for $z\,>0$ the scattering amplitude shows the geometric scaling behaviour being the function only one variable: $z$\cite{BALE,GS}.
 
 Fourth, for $z \,\gg\,1$ the solution to BK has been found\cite{LETU}:
 \beq \label{LTSOL}
  N_{01}\Lb  z\Rb \,\,=\,\,1 \,\,-\,\,C(z) \exp\Lb - \frac{z^2}{2\,\kappa}\Rb
  \eeq
 where $C(z)$ is a smooth function of $z$. 
  
 Using \eq{SIMKER} we can evaluate $\gamma_{cr} = \h$ and  $\kappa = 4$.  Solution in the vicinity of the saturation scale we have found in the previous subsection (see \eq{RH12}). One can compare it with \eq{VQS}. However, for  our simplified kernel we can find the solution to BK equation in the entire saturation region: $z>0$.

Inside the saturation region the BK equation  with the kernel of \eq{K2}  takes the form
\beq \label{BK1}
\frac{\partial^2 \widehat{N}\Lb Y, \xi; \vec{b}\Rb}
{ \partial Y\,\partial \xi}\,\,=\,\, \bas \,\left\{ \Lb 1 
\,\,-\,\frac{\partial \widehat{N}\Lb Y, \xi; \vec{b}
 \Rb}{\partial  \xi}\Rb \, \widehat{N}\Lb Y, \xi;
 \vec{b}\Rb\right\}
\eeq
where 
 $\widehat{N}\Lb Y, \xi; \vec{b}\Rb\,\,=\,\,\int^{\xi} d \xi'\,N\Lb Y,
 \xi'; \vec{b}\Rb$ . Introducing 
     $\Omega\Lb Y; \xi, \vec{b}\Rb$\cite{LETU}
\beq \label{SOL1}
N\Lb Y, \xi \Rb\,\,=\,\,1\,\,-\,\,\exp\Lb - \Omega\Lb Y, \xi\Rb\Rb
\eeq
we can rewrite \eq{BK1}  as
\begin{subequations}
\bea \label{SOL2}
&&\frac{ \partial \Omega\Lb Y, \xi\Rb}{ \partial Y} \,\,=\,\,\bas \widetilde{N}\Lb Y, \xi\Rb;~~~\frac{ \partial^2 \Omega\Lb Y, \xi\Rb}{ \partial Y\,\partial \xi}  \,\,\,=\,\,\bas \Bigg( 1 -\,\exp\Lb - \Omega\Lb Y, \xi\Rb\Rb\Bigg);\label{SOL02}\\
&&~\frac{ \partial^2 \Omega\Lb \xi_s; \zeta\Rb}{ \partial \xi_s\,\partial \xi}  \,\,\,=\,\,\frac{1}{4}\Bigg( 1 -\,\exp\Lb - \Omega\Lb \xi_s; \zeta\Rb\Rb\Bigg)
\eea
\end{subequations}
The variable $\xi_s$ is
 defined as 
\beq \label{XISZ}
\xi_s\,\,=\,\,\ln\Lb Q^2_s\Lb Y\Rb/Q^2_s\Lb Y=0; \vec{b},\vec{R}\Rb\Rb\,\,=\,\,4 \bas\,Y;~~~~~~z\,\,=\,\,\xi_s\,\,+\,\,\xi
\eeq
\eq{SOL2} has a traveling wave solution (see formula {\bf 3.4.1.1} of
  Ref.\cite{MATH}). For \eq{SOL2} in the canonical form:
\beq \label{SOL3}
\frac{ \partial^2 \Omega\Lb \xi_s; \tilde{ \zeta}\Rb}{ \partial t^2_+} \,\,-\,\,\frac{ \partial^2 \Omega\Lb \xi_s; \tilde{\zeta}\Rb}{ \partial t^2_-} \,\,\,=\,\,\frac{1}{4}\,\,\Bigg( 1 -\,\exp\Lb - \Omega\Lb \xi_s; \tilde{\zeta} \Rb\Rb\Bigg),
\eeq
with $t_{\pm} = \xi_s \pm  \xi $, the solution takes the form:
\beq \label{SOL4}
\int^{\Omega}_{\Omega_0}\frac{d \Omega'}{\sqrt{ C_1 + \frac{2}{(\mu^2 - \kappa^2)} \frac{1}{4} \Lb \Omega' + \exp\Lb-\Omega'\Rb\Rb}}\,\,=\,\,\mu t_+ + \kappa t_-  + C_2
\eeq
where all constants have to be determined  from the initial and boundary 
conditions of \eq{RH12}. First we see that $C_2=0 $ and $\kappa =0$.
 From the condition $\Omega'_z/\Omega \,\,=\,\,\h$ at $t_+ =
 0$ we can find $C_1$. Indeed, differentiating \eq{SOL4} with respect to 
$t_+$ one
 can see that at $t_+= 0$ we have:
\beq \label{SOLSET}
\frac{d \Omega}{d t_+}|_{t_+ = 0}\,\frac{1}{\sqrt{ C_1 \,\,+\,\,\frac{1}{2\,\mu^2}\Lb 1\,\,+\,\,\h\,\Omega^2_0\Rb}}\,\,=\,\,\mu
\eeq
From     \eq{SOLSET} one can see that choosing 
\beq \label{SOLSET1}
C_1\,\,=\,\,- \,2;~~~~~\mbox{and}\,~~~\mu \,\,=\,\,\h 
\eeq
we satisfy the initial condition $\frac{d \ln\Lb \Omega\Rb}{d z}|_{t_+ = 0}
 =\,h$ of \eq{VQS} or \eq{RH12} for our case.  At large $z$ we obtain the solution\cite{LETU}:
\beq \label{SOL8}
\Omega\Lb z\Rb\,\,=\,\,\frac{1}{8}\,z^2\,\,+\,\,{\rm Const};
~~~~~
N\Lb z\Rb\,\,=\,\,1\,\,-\,\,{\rm Const}\,e^{-\,\frac{1}{8}\,z^2};
\eeq
  We wish  to stress that  \eq{SOL8} reproduces the 
 asymptotic solution to \eq{BK1}, which has been derived in
 Ref.\cite{LETU}, for fixed $\bas$.  
Finally, the solution of \eq{SOL4} can be re-written in the following
 form for $\Omega_0 \ll 1$:
\beq \label{SOL5}
\frac{1}{\sqrt{2}}\int^{\Omega}_{\Omega_0}\frac{d \Omega'}{\sqrt{\Omega'\,\,+\,\,\,e^{ - \,\Omega'}\,\,-\,\,1
}}\,\,=\,\, \h\,z \eeq
\eq{SOL5} gives the solution which depends only on one variable
 $z\,\,=\,\,\xi_s + \xi$,
and satisfies the initial conditions of \eq{VQS}.  

Therefore, we have shown that our simplified version of the BFKL kernel describes  all general feature of the
BK equation and give the analytical solution to it.

~

~

     \begin{boldmath}
     \subsection{
   Equations for parton densities. }
      \end{boldmath}

The equations for the parton densities $\rho_n$ have been discussed in Ref.\cite{LE1}. Here we    introduced them just for a completeness of presentation.
We can introduce the parton densities  $\rho_n(r_1, b_1,\ldots\,,r_n, b_n)$ using the generating functional $Z$:
\beq \label{PD}
\rho_n(r_1, b_1\,
\ldots\,,r_n, b_n; Y\,-\,Y_0)\,=\,\frac{1}{n!}\,\prod^n_{i =1}
\,\frac{\delta}{\delta
u_i } \,Z\left(Y\,-\,Y_0;\,[u] \right)|_{u=1}
\eeq
where $Z$ is defined as\cite{MUDI}
\beq \label{Z}
Z\Lb Y, \vec{r},\vec{b}; [u_i]\Rb\,\,=\,\,\sum^{\infty}_{n=1}\int P_n\Lb Y,\vec{r},\vec{b};\{\vec{r}_i\,\vec{b}_i\}\Rb \prod^{n}_{i=1} u\Lb \vec{r}_i\,\vec{b}_i\Rb\,d^2 r_i\,d^2 b_i
\eeq
 where $u\Lb \vec{r}_i\,\vec{b}_i\Rb \equiv\,u_i$ is an arbitrary function and $P_n\Lb Y, r, b ; \{r_i,b_i\}\Rb$ is the probability to have $n$-dipoles
 of size $r_i$,  at impact parameter $b_i$ and  at rapidity $Y$.  
The initial and  boundary conditions for \eq{Z}  take 
the following form for the functional $Z$:
\begin{subequations}
\bea
Z\Lb Y=0, \vec{r},\vec{b}; [u_i]\Rb &\,\,=\,\,&u\Lb \vec{r},\vec{b}\Rb;\label{ZIC}\\
Z\Lb Y, r,[u_i=1]\Rb &=& 1; \label{ZSR}
\eea
\end{subequations}
In QCD we have the following equations for $P_n$:
    \beq  \label{PC1}
\frac{\partial\,P_n\Lb Y, \vec{r }, \vec{b};\{\vec{r}_i,\vec{ b}_i\} \Rb}{ 
\partial\, Y }\,=\,-\,
\sum^n_{i=1}\,\omega_G(r_i) \,
P_n\Lb Y, \vec{r }, \vec{b};\,\{\vec{r}_i,\vec{ b}_i\},\Rb \,\,+\,\,\bas\,\sum^{n-1}_{i=1} \,\frac{(\vec{r}_i\,+\, 
\vec{r}_n)^2}{(2\,\pi)\,r^2_i\,r^2_n}\,
P_{n - 1}\Lb Y, \vec{r},\vec{b};\{\vec{r}_j, \vec{b}_j ,\vec{r}_i+ \vec{r}_n,\vec{b}_{in}\}
\Rb\nn
\eeq
   \eq{PC1} is a typical cascade equation in which the first term
 describes the reduction   of  the probability to find $n$ dipoles
 due to the possibility that one of $n$ dipoles can decay into two dipoles 
of
 arbitrary sizes  
  , while the second term,  describes  the growth due to the 
splitting
 of $(n-1)$ dipoles into $n$ dipoles.   

From \eq{PC1} in Refs.\cite{LELU,LE1} the set of equations for  $\bar{\rho}_n(\rv, \bv; \{r_i, b_i\})$
defined as
 \beq \label{BRHO}
\bar{\rho}_n(\rv, \bv; \{r_i, b_i\}) \,=\,\,\prod_{i=1}^n \,r^2_i\,\,\rho_n(\{r_i, b_i\})
\eeq has been derived:
\bea \label{PD3}
\frac{\partial \,\bar{\rho}_n(\{\rv_i, \bv_i\})}{ 
\,\partial\,Y}\,\,&=&\,\,\sum_{i=1}^n\,
\int\,\frac{d^2\,r'}{2\,\pi}\,
K\Lb \vec{r}',\vec{r}_i - \vec{r'}|\vec{r}_i\Rb\\
&\times&\Bigg\{ \bar\rho_n(\{\rv_j,\bv_j\}, \rv',\bv_i - (\rv_i - \rv')/2)
\,+\,\bar\rho_n(\{\rv_j,\bv_j\}, \rv_i -  \rv',\bv_i - \rv'/2)\,-\,\bar{\rho}_n(\{\rv_i, \bv_i\})\Bigg\}
\nn\\
 & & 
\,
+\,\bas\sum_{i=1}^{n-1}\,
\bar\rho_{n-1}(\ldots\,(\vec{r}_i\,+\,\vec{r}_n), b_{in}\dots).\nn
\eea
It turns  out that for the initial condition of \eq{ZIC} we 
can obtain  the nonlinear equations for  $\bar{\rho}_n(\{\rv_i, \bv_i\})$ \cite{LE1}:
 \bea \label{RHONNEQ}
 &&\frac{\partial \,\bar\rho_n(\rv, \bv; \{\rv_i, \bv_i\})}{ 
\,\partial\,Y}\,\,=\,\, \bas\,\int d^2 r'\,  K\Lb \vec{r}',\vec{r} - \vec{r'}|\vec{r}\Rb\\
& \times&\,\,\Bigg\{\Lb\bar\rho_n(\rv', \bv- \h(\rv - \rv'); \{\rv_i, \bv_i\}) \,\,+\,\,\bar\rho_n(\rv - \rv', \bv- \h\rv'; \{\rv_i, \bv_i\}) -\,\,\bar\rho_n(\rv, \bv; \{\rv_i, \bv_i\})\Rb\nn\\
 &+&\bas\,\,\sum^{n-1}_{k=1} \bar\rho_{n-k}(\rv', \vec{b} + \h(\vec{r} - \vec{r}'; \{\rv_i, \bv_i\}) \,\bar\rho_k(\rv -\rv', \bv - \h \rv'; \{\rv_i, \bv_i \})\Bigg\} \nn
 \eea
This equation together with \eq{PD3} leads to the recurrence relation\cite{LE1} for $\bar \rho_n$, which has the form:
 \bea \label{PD4}
&&\int d^2 r'\,  K\Lb \vec{r}',\vec{r} - \vec{r'}|\vec{r}\Rb\,\Bigg\{\Lb\bar\rho_n(\rv', \bv- \h(\rv - \rv'); \{\rv_i, \bv_i\}) \,\,+\,\,\bar\rho_n(\rv - \rv', \bv- \h\rv'; \{\rv_i, \bv_i\}) -\,\,\bar\rho_n(\rv, \bv; \{\rv_i, \bv_i\})\Rb\nn\\
 &+&\,\sum^{n-1}_{k=1} \bar\rho_{n-k}(\rv', \vec{b} + \h(\vec{r} - \vec{r}'; \{\rv_i, \bv_i\}) \,\bar\rho_k(\rv -\rv', \bv - \h \rv'; \{\rv_i, \bv_i\}) \Bigg\}\,\,=\,\,
  \sum_{i=1}^n\,
\int\,\frac{d^2\,r'}{2\,\pi}\,
K\Lb \vec{r}',\vec{r}_i - \vec{r'}|\vec{r}_i\Rb \nn\\
&\times&\Bigg\{ \bar\rho_n(\rv, \{\rv_j,\bv_j\}, \rv',\bv_i - (\rv_i - \rv')/2)
\,+\,\bar\rho_n(\rv,\{\rv_j,\bv_j\}, \rv_i -  \rv',\bv_i - \rv'/2)\,-\, \bar{\rho}_k \bar{\rho}_{n - k}\Bigg\}
\nn\\
 &+ & 
\,
\,\,\sum_{i=1}^{n-1}\,
\bar\rho_{n-1}(\rv, \ldots\,(\vec{r}_i\,+\,\vec{r}_n), b_{in}\dots)
\eea
 

    ~

     \begin{boldmath}
     \subsection{ Solution for $\bar{\rho}_2\Lb Y, r, r_1,b_1, r_2,b_2\Rb$}
      \end{boldmath}     


    ~

   The recurrence relation of \eq{PD4} for  $ \bar{\rho}_2(Y, \xi_1, \xi_2)  $   has the following form for the kernel of \eq{K2}:
    
   \beq \label{RH2}    
   \int\limits^{\zeta_{2}}_0\!\!d  \zeta_2'  \bar{\rho}_2(\zeta_2')\,\,+\,\,\h\,\Big( \int\limits^{z_1}_0 \!\!d z_1'  \rho_1\Lb z_1'\Rb \,\rho_1\Lb z_2\Rb\,\,+\,\, \int\limits^{z_2}_0\!\! d z_2'  \rho_1\Lb z_1\Rb \,\rho_1\Lb z_2'\Rb \Big) = \,\,\int\limits^{z_1}_0 d  z_1'   \bar{\rho}_2(z_1' , z_2)\,\,+\,\,
  \int\limits^{z_2}_0 d  z_2'   \bar{\rho}_2(z_1 , z_2') 
 \eeq    
with $\zeta_2 = \h\Lb z_1 + z_2\Rb$. In \eq{RH2} we use  \eq{K2} for the leading twist BFKL kernel and 
\beq \label{sumxi}
\h\Lb \xi'_1 + \xi'_2\Rb = \ln r^2 \,+\,\h\ln\Lb\frac{r^2_1\,r^2_2}{b^4_1\,b^4_2}\Rb
\eeq
Implicitly we assume that $r_1 \sim r_2, b_1 \sim b_2$ . We will see below that integration over $r_1$ and $r_2$ as well as over $b_1$ and $b_2$ will be independent in \eq{MPSI} and they  lead to $\frac{ r^2\,r_1^2}{b_1^4}
\exp\Lb 4\bas\,Y\Rb \,\approx \,1$ and $ \frac{ r^2\,r_2^2}{b_2^4}
\exp\Lb 4\bas\,Y\Rb \,\approx \,1$.
.   
   One can check that 
   \beq \label{RH21}     
   \bar{\rho}_2(z_1 , z_2)  \,\xrightarrow{z_1 \gg1, z_2 \gg 1}\,\,\frac{2}{3}\rho_1\Lb z_1\Rb  \,\rho_1\Lb z_2\Rb     \eeq
   is the solution of \eq{RH2}.

   \eq{RH21} support an assumption that 
     \beq \label{RHN}  
  \rho_n\Lb \zeta_n\Rb\,\,=\,\,{\rm C_n}\,\prod_{i=1}^n \rho_1\Lb z_i\Rb
    \eeq
      for $z_i \gg 1$ . $\zeta_n = \frac{1}{n} \sum_{i=1}^n z_i$.

      ~

      ~
     \begin{boldmath}
     \subsection{ Solution for $\bar{\rho}_n\Lb Y, r, \{r_1,b_i\}\Rb$}
      \end{boldmath}     
        Actually, it turns out that it is more convenient to use  \eq{RHONNEQ}  for proving \eq{RHN}. This euation  for the kernel of \eq{K2} takes the form:
      \beq \label{RHN1}
        \Lb  4\frac{ d^2}{d \zeta_n^2} \,\,-\,\,1\Rb\,\rho_n\Lb \zeta_n\Rb\,=\,\,\h\frac{d}{d \zeta_n} \sum_{ i=1}^n \Big( \int^{\zeta_{n-i}}_0\!\!\! \!\!\! d \zeta_{n-i}'\,\, \rho_{n-i}\Lb z_{n-i}'\Rb \,\,\rho_i\Lb \zeta_i\Rb \,\,+\,\, \rho_{n-i}\Lb z_{n-i}'\Rb \,\,\int^{\zeta_{i}}_0 \!\!\!d \zeta_{i}'\,\,\rho_i\Lb \zeta_i'\Rb \Big)
        \eeq         
    Plugging \eq{RHN} in this equation one can see that for ${\rm C_n}$ we obtain the following equation:

        \beq \label{RHN2}      
    \Lb  n^2-1 \Rb\,{\rm C_n} \,\,=\,\,\frac{n^2}{2} \sum_{i=1}^{n-1} \frac{ {\rm C_{n-i}\,C_i}}{(n-i)\,i}
      \eeq
        
     The first ${\rm C_n}$  are  1,2/3,3/8 for $n= 1,2,3$ and ${\rm C_n} = e^{- \alpha \,n} 2 \,n$ for $n \gg 1$., with arbitrary parameter $\alpha$.  $\alpha$ could be found from matching with ${ \rm C_1}$ for example.

       \eq{RHN2} we can solve introducing the generating function $\cal Z$:
       \beq \label{LT1}
     {\cal  Z}\Lb u\Rb\,\,=\,\,\sum^\infty_{n=1} \frac{\rm C_n }{n} u^n  
     \eeq

       The initial conditions for $Z(u)$ have the following form:
       \beq \label{LT2}
     {\cal  Z}\Lb u = 0\Rb = 0;~~~~~\frac{d \,{\cal Z}\Lb u \Rb}{ d\,u}\Big{|}_{u=0} = 1;
      \eeq
        
      \eq{RHN2}   can be rewritten as the following equation for $\cal Z(u)$:
      
      \beq \label{LT3}
        u\,\frac{d \,{\cal Z}(u)}{d u}  - \int^u_0\frac{{\cal Z}\Lb u'\Rb}{u'} d u'\,\,=\,\,\h{\cal Z}^2\Lb u\Rb~~~~\mbox{or}~~~ u \frac{d}{d\,u} u \frac{d}{d\,u} {\cal Z}\Lb u \Rb \,\,-{\cal Z}\Lb u\Rb\,\,=\,\,\h u \frac{d}{ d\,u} {\cal Z}^2\Lb u\Rb
        \eeq

        Introducing a new variable $\xi = \ln u$ and      
        assuming that $ \frac{d \,{\cal Z}\Lb \xi\Rb }{ d \,\xi} = p({\cal Z}) $ we obtain:
        
        \beq \label{LT4}
       p({\cal Z}) \frac{d p({\cal Z})}{d {\cal Z}} =  {\cal Z} \Lb 1 \,+\, p({\cal Z})\Rb
        \eeq
        with the solution:
        \beq \label{LT5}
        p({\cal Z}) - \ln\Lb 1 + p({\cal Z})\Rb\,\,=\,\, \h {\cal Z}^2 + \kappa_1
        \eeq
        where $\kappa_1$ is  a constant, From   \eq{LT2}  it  is equal to zero.  Using variable $\xi$ we can 
        calculate the coefficients ${\rm C_n}$ . From \eq{LT1} one can see that
        \beq \label{LT6}
          \frac{\rm C_n }{n}\,\,=\,\,\,\oint\frac{ d\,u}{2\,\pi\,i}  \,\frac{ {\cal Z}\Lb u\Rb}{u^{n+1}} \,\,=\,\,\oint\frac{ d\,\xi}{2\,\pi\,i}  e^{ - \xi\,n} \, {\cal Z}\Lb \xi \Rb \,\,= \, \,\int^{2 \pi}_0\frac{ d \phi}{2 \pi}e^{ - i\,\phi\,\,n} {\cal Z} \Lb \phi \Rb \eeq
          
          The contour of integration is the unit circle around $u = 0$ and in the last equation we introduce $\xi = i \,\phi$. We use the notation ${\cal Z}\Lb \phi\Rb$ instead of ${\cal Z}\Lb e^{i\,\phi}\Rb$.
          
      It turns out that \eq{LT6} is useful for estimates of the scattering amplitude. Two limits:  small $Z(u)$ and 
      large ${\cal Z}\Lb u \Rb$, have simple analytical solutions. For $ Z \ll  1$ \eq{LT5} reads:
      \beq \label{LT7}
      \h p^2= \h {\cal Z}^2~~~~\mbox{leading \,to} \frac{ d{\cal Z}}{d \xi} = {\cal Z}      \eeq
      with solution:
      \beq \label{LT8}
      {\cal Z}\Lb \xi\Rb= \kappa e^{ \xi}
      \eeq 
      with constant $\kappa=1$ from \eq{LT2}.
      
      For ${\cal Z}\Lb \xi \Rb\,\,\gg\,1$  we have from \eq{LT5}
       \beq \label{LT9}
       p= \h {\cal Z}^2~~~~\mbox{leading \,to} \frac{ d {\cal Z}}{d \xi} = \h\,{\cal Z}^2
      \eeq      
      The solution to \eq{LT9} takes the form:
      \beq \label{LT10}
     {\cal Z}\Lb \xi\Rb\,\,=\,\,\frac{2}{\xi \,-\,\kappa_2} 
         \eeq

          Using \eq{LT6} one can obtain that 
          \beq \label{LT11}
        {\rm C_n} \,\,=\,\,2\,n e^{- \kappa_2 n} 
          \eeq
          reproducing the solution of \eq{RHN2} at large $n$.
          
\begin{figure}[ht]
    \centering
 \leavevmode
 \begin{tabular}{c c} 
    \includegraphics[width=8cm]{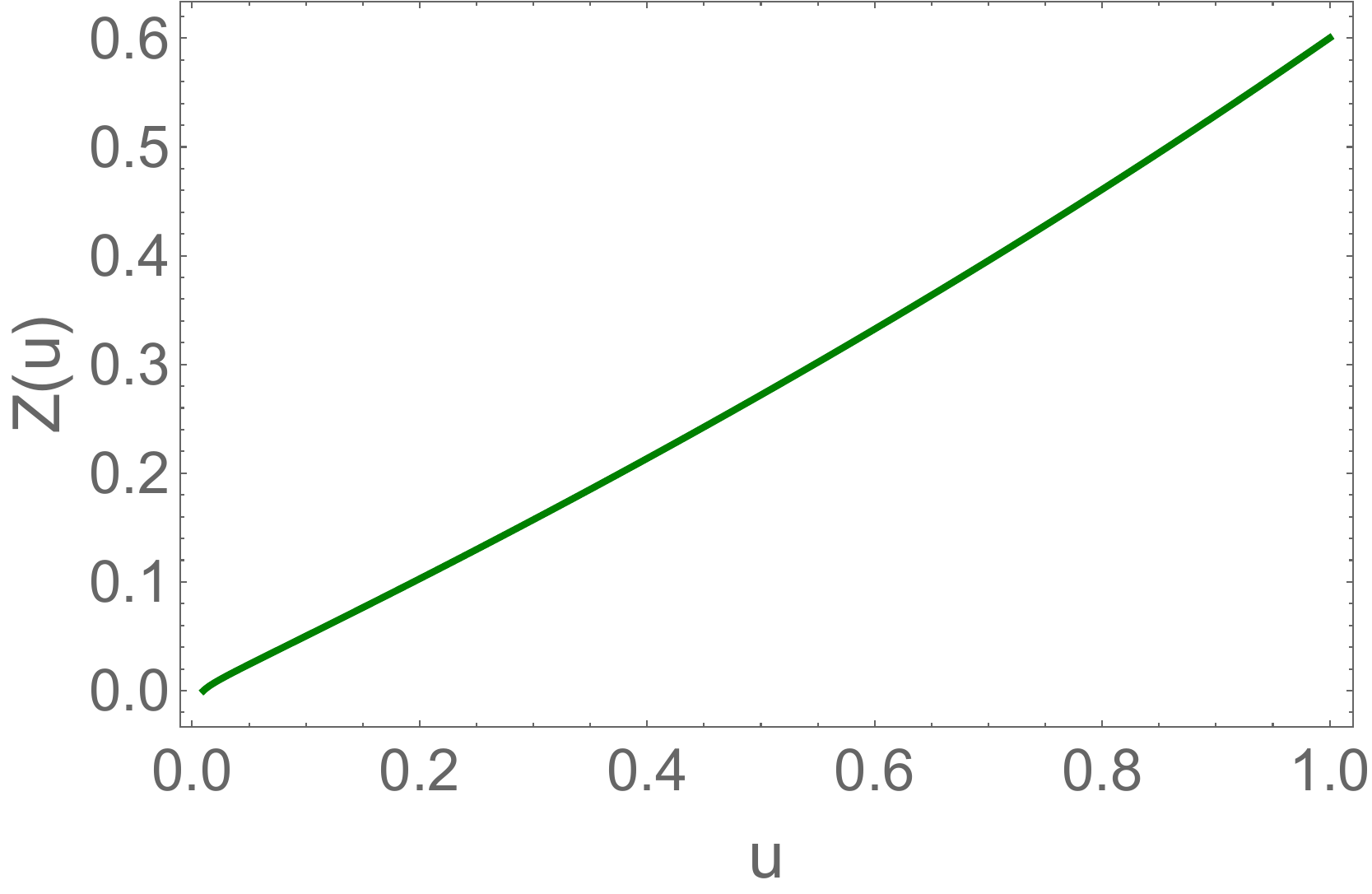}  &\includegraphics[width=8.35cm]{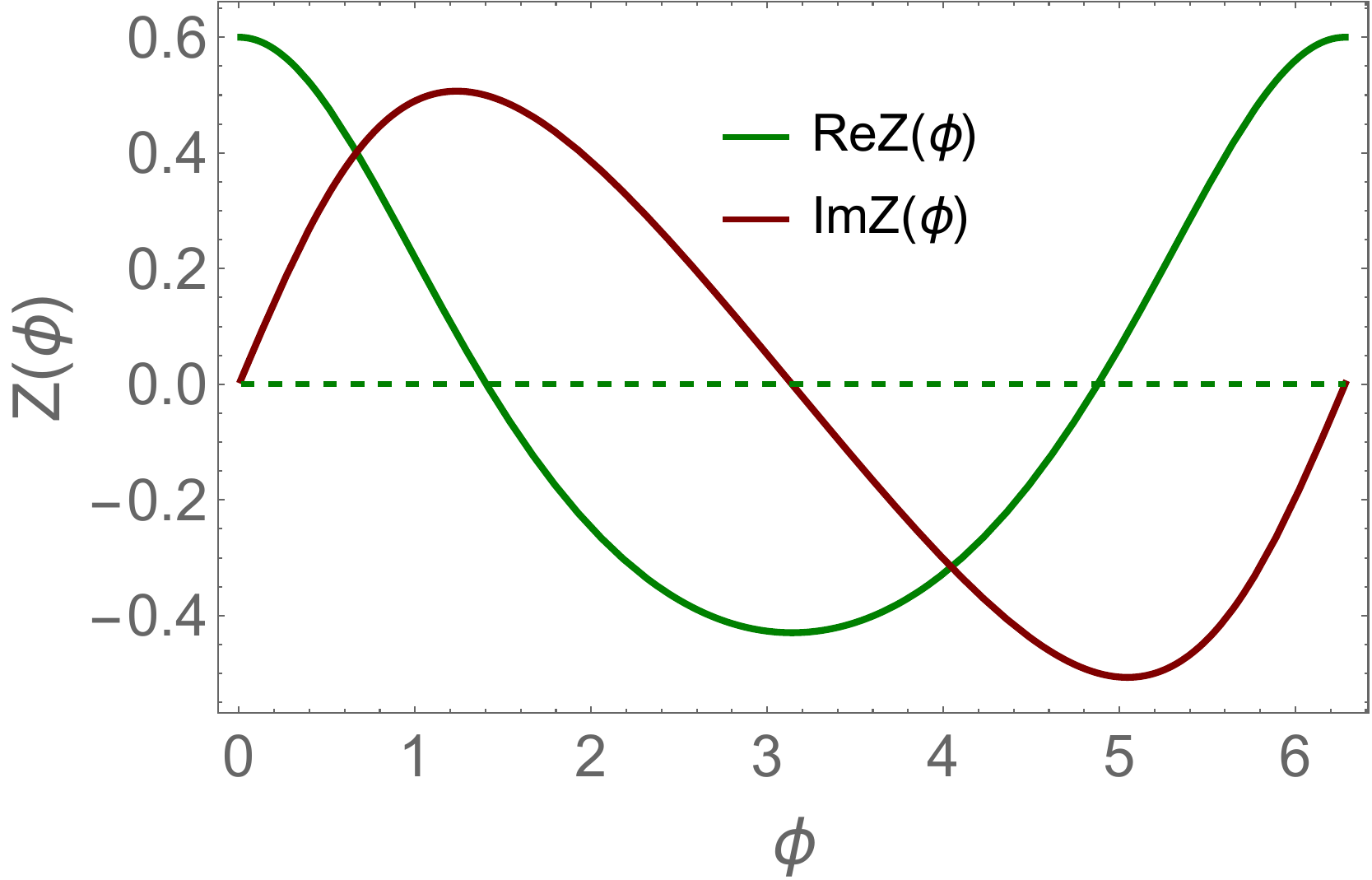} \\
    \end{tabular}
        \caption{\fig{sol}-a:Solution to \eq{LT3}  with the initial conditions of \eq{LT2}. \fig{sol}-b: Solution to \eq{LT12} with the initial condition of \eq{LT13}.}
\label{sol}
   \end{figure}

          As  will  be seen below, we need to know $C_n$ at all values of $n$.  Unfortunately, we failed to find an analytical   solution   for all $n$. Hence, we have to approach the problem using numerical solutions. We proceed solving two equations.  First we solve  \eq{LT3} with the initial conditions of \eq{LT2} and using this solution we fix the initial condition for the generation functional at $u=1$. The solution is shown in \fig{sol}-a.    One can see that from this solution we obtain:
           \beq \label{LT111} 
   {\cal Z}\Lb u=1\Rb = 0.6;~~~~~\frac{d \,{\cal Z}\Lb \phi \Rb}{ d \,u}\Big{|}_{u=1} = \,0.72;          
            \eeq    
          Second, rewriting \eq{LT3} for $\xi = i\,\phi$ we obtain that
          
  \beq \label{LT12}
    \frac{d^2}{d\,\phi^2} {\cal Z}\Lb \phi \Rb \,\,+\,\,{\cal Z}\Lb \phi\Rb\,\,=\,\,i\,{\cal Z}\Lb \phi\Rb \frac{d}{ d\,\phi} {\cal Z}\Lb \phi\Rb
        \eeq
    with the initial conditions of \eq{LT111} . In the variable $\phi$ they have the form:
     \beq \label{LT13}
{\cal Z}\Lb \phi=0\Rb = 0.6;~~~~~\frac{d \,{\cal Z}\Lb \phi \Rb}{ d\,\phi}\Big{|}_{\phi=0} = i \,0.72;
      \eeq       
        
        The solution is plotted in \fig{sol}-b.  
                   
              ~
%
             ~          ~

          ~

        ~
      
     \begin{boldmath}
     \section{ Scattering amplitude}
      \end{boldmath}

      
        As it has been mentioned, our main goal is
to sum large Pomeron loops in QCD to obtain the scattering amplitude. In our previous paper \cite{LE1} 
 we showed that the scattering amplitude in the Pomeron calculus in two dimension which satisfies both $s$ and $t$ channel unitarity, can be reproduced in the BFKL cascade.  The amplitude  in this model can be viewed as the sum of large Pomeron loops as it is illustrated in \fig{mpsi}-b. Having this result we conjecture that the sum of the large BFKL Pomeron loops  determine the scattering amplitude at high energies.
As we have mentioned in the introduction 

 Actually, the problem of summation of the large Pomeron loops in QCD 
 has been solved and it is given by \eq{MPSI}.

  For 
 \eq{RHN}  we can simplify \eq{MPSI} using the  $t$-channel unitarity constraints for the BFKL Pomeron\footnote{Actually the BFKL Pomeron is derived using the $t$-channel unitarity\cite{BFKL}. In Ref.\cite{MUSA}  the $t$-channel unitarity is rewritten in coordinate representation.} , which takes the  following form in the coordinate representation \cite{CLMSOFT}:
 \beq \label{BFKLT}
  G^{\rm BFKL} \Lb Y, r,R ;  \vec{b}\Rb\,\,=\,\,\frac{1}{4\,\pi^2}\int \frac{d^2 r_i}{r^4_i} \,d^2 b_i 
   \,   G^{\rm BFKL} \Lb Y - Y_0, r, r_i;  \vec{b}\ - \vec{b}_i \Rb \,\, G^{\rm BFKL} \Lb Y_0,  r_i;  \ \vec{b}_i \Rb  
   \eeq  
   
   Using \eq{BFKLT} we
  perform the integration over $r'$    and the  final formula for the scattering amplitude takes the form:
         \beq \label{MPSI1}
      A\Lb Y, r, R ;  \vec{b}\Rb\,=\,\sum^\infty_{n=1}\,\Lb -1\Rb^{n+1}\,n!\int  \prod \frac{d^2 r_i}{4\, \pi^2\,r^4_i}\,\,d^2 b'_i 
     \int d^2 \delta b_i     \,\,\bar{\rho}\Lb Y - Y_0, \{ \vec{r}_i,\vec{b}_i\}\Rb\,\bar{\rho}\Lb Y_0, \{ \vec{r}_i,\vec{b}'_i\}\Rb   \eeq

    ~

  
     \subsubsection{The BFKL Pomeron exchange}

    The first term of \eq{MPSI1} is 
     the contribution of the single  BFKL Pomeron exchange to the scattering amplitude which takes the form\cite{MUDI}:
    \beq \label{BFKL1}
    A^{\mbox{\tiny BFKL}}\Lb Y, \vec{r}; \vec{R},  \vec{b}\Rb\,\,=\,\,\int \frac{d^2\,r_1}{r^4_1}   \,d^2 b_1\,\bar\rho\Lb  Y  - Y_0, \vec{r};  \rv_1, \vec{b} - \vec{b}_1\Rb \,\,  \,\, \bar\rho\Lb   Y_0, \vec{R};  \rv_1,  \vec{b}_1\Rb 
    \eeq
 
 \eq{RH12} can be written using \eq{EIGENF},  as
 \beq \label{BFKL2}
 \rho_1\Lb z\Rb\,\,=\,\,N_0 \Lb \frac{e^{\lambda\,\bas\,Y} r^2\,r_1^2}{\Lb \vec{b}   + \h\vec{r} \Rb^2\,\Lb \vec{b}  -  \h\vec{r} \Rb^2}\Rb^{1/2} 
 \eeq
 In \eq{BFKL2} we took into account that $r_1 \ll b$  for our BFKL kernel ,
 
 Plugging this equation  into  \eq{BFKL1} we obtain
   \beq \label{BFKL3}
  A^{\mbox{\tiny BFKL}}\Lb Y, \vec{r}; \vec{R},  \vec{b}\Rb = 
 N^2_0 \intl \!\! \frac{d^2\,r_1}{r^4_1}   \,d^2 b_1\Lb \frac{e^{\lambda\,\bas\,\Lb Y - Y_0\Rb} r^2\,r_1^2}{\Lb \vec{b} - \vec{b}_1  + \h\vec{r} \Rb^2\,\Lb \vec{b} - \vec{b}_1  -  \h\vec{r} \Rb^2}\Rb^{1/2} \,  \Lb \frac{ e^{\lambda\,\bas\,Y_0}R^2\,r_1^2}{\Lb \vec{b}_1  + \h\vec{R} \Rb^2\,\Lb \vec{b}_1  -  \h\vec{R} \Rb^2}\Rb^{1/2}  \eeq 
   
  Recall, that  in our region of integration  for $r_1$ and $b_1$ both parentheses are larger 1. First : one can see that \eq{BFKL3}  that the main dependence on $Y_0$ disappear.  Second, we see two log integrals in \eq{BFKL3}: over $r_1$ and $b_1$.   The log integral over $b_i$ stems from $b\,\gg\,b_1 \,\gg\,R$ and, therefore, sufficiently large $b$ are essential for high energy scattering.   All integration can be done in a general form if we remember that  \eq{RH12}   is the solution to the linear BFKL equation with the leading twist kerner in the vicinity of the saturation scale. Hence, we can rewrite $ A^{\mbox{\tiny BFKL}}\Lb Y, \vec{r}; \vec{R},  \vec{b}\Rb$ in the form:
     \bea \label{BFKL31}
  &&A^{\mbox{\tiny BFKL}}\Lb Y, \vec{r}; \vec{R},  \vec{b}\Rb =  N^2_0\intl^{i\epsilon +  \infty}_{i\epsilon - \infty}\!\!\!\frac{ d \nu_1}{2\,\pi\,i} \intl^{i\epsilon +  \infty}_{i\epsilon - \infty}\!\!\!\frac{ d \nu_2}{2\,\pi} \intl \!\! \frac{d^2\,r_1}{r^4_1}   \,d^2 b_1 \nu_1\,\nu_2\,\,e^{\bas \Lb \chi\Lb \nu_1\Rb \Lb Y - Y_0\Rb\,+\,\chi\Lb \nu_2\Rb\,Y_0\Rb}\nn\\
&&\times\,\,  
\Lb \frac{ r^2\,r_1^2}{\Lb \vec{b} - \vec{b}_1  + \h\vec{r} \Rb^2\,\Lb \vec{b} - \vec{b}_1  -  \h\vec{r} \Rb^2}\Rb^{1/2 + i \nu_1} \,  \Lb \frac{ R^2\,r_1^2}{\Lb \vec{b}_1  + \h\vec{R} \Rb^2\,\Lb \vec{b}_1  -  \h\vec{R} \Rb^2}\Rb^{1/2 - i \nu_2}  \eea 
Note, that the contributions of the BFKL Pomerons  have extra factors $\nu_1$ and $\nu_2$ which provide the correct behaviour in the vicinity of the saturation scale ( see Refs.\cite{LIP,MUDI,MUMU}).

  Integration over $r_1$ leads to $\nu_1 = \nu_2$ while logarithmic integration over $b_1$ gives  the following expression:
    \beq \label{BFKLNK}
A^{\mbox{\tiny BFKL}}\Lb Y, \vec{r}; \vec{R},  \vec{b}\Rb = 
 N^2_0\intl^{\epsilon + i \infty}_{\epsilon - i \infty}\!\!\!\frac{ \nu_1 \,d \nu_1}{2\,\pi}  e^{\bas  \chi_{\rm LT}\Lb \h + i\nu_1\Rb Y  } \Lb\frac{ r^2\,R^2}{\Lb \vec{b} + \h\vec{r} \Rb^2\,\Lb \vec{b} -  \h\vec{r} \Rb^2}\Rb^{1/2 + i \nu_1}    
  \,\,=\,\,\tilde{N}_0 e^{ \h z} \,z
  \eeq
  with
    \beq \label{BFKL4}
    e^{ z}   \,\,=\,\,   \frac{e^{\lambda\,\bas\,Y} \,r^2\,R^2}{b^4} 
    \eeq   
  In \eq{BFKL31} we collect all constants in $\tilde{N}_0 $ and denote by  $\chi_{\rm LT}$  the BFKL kernel in the leading twist approximation. Recall, that $\lambda$  in \eq{BFKL4} is equal to $ \chi_{\rm LT}\Lb \bar{\gamma} \Rb /\bar{\gamma} $ with $\bar {\gamma} = \h$.

    From \eq{BFKL3} and \eq{BFKL31} one can see that the typical values of $b$ are  large  spanning from $R \approx r $ to  $b$ with maximal $b =  \Lb e^{\lambda\,\bas\,Y} \,r^2\,R^2  \Rb^{1/4}$.  Thus we confirm the results of Refs.\cite{MUDI,MUSA} that large $b$ contribute to the scattering amplitude.  It has been  obtained in  the diffusion approximation to the BFKL kernel, and we show that it  is also correct  for the leading twist BFKL kernel. Bearing this in mind we could could consider $\delta b_i$ in  \eq{MPSI} 
 as being small and  replace  
 
 \beq \label{BFKL5}  
 \int d^2 \delta b_i \gamma^{BA}\Lb r_1,r'_i, \vec{b}_i -  \vec{b'_i}\equiv \delta \vec{b} _i\Rb \,\,=\,\,\sigma^{BA} \Lb r_i, r'_i \Rb\,=\,4 \bas^2\,\int\frac{d \,l}{l^3} \Lb 1 - J_0\Lb l\,r_i\Rb\Rb\,\Lb 1 - J_0\Lb l\,r'_i\Rb\Rb
 \eeq
    Using \eq{BFKL5} we can calculate the BFKL contribution to the scattering amplitude directly from \eq{MPSI}.   Doing so we obtain the same result as \eq{BFKLNK} with a different constant in front.  
This difference stems from different initial conditions for $\rho_1$ in \eq{MPSI} and BFKl  Pomeron Green's function    in \eq{BFKLT}. and can be absorbed in $N_0$, which is not deremined in our approach and does not influence on any of the results.

     \begin{boldmath}
     \subsubsection{Summing large Pomeon loops   for the BFKL leading twist kernel}
      \end{boldmath}

    We can estimate the scattering amplitude of two dipoles 
   using \eq{MPSI}  and \eq{BFKLNK}. It results in the following equation for S-matrix , $ S = 1 - A$:
   
     \bea \label{MPSI11}
S\Lb Y, \rv,\vec{R} ;  \vec{b}\Rb\,=\,S\Lb z\Rb&=&\,\sum^\infty_{n=0}\,\Lb -1\Rb^{n}\,n!\,{\rm C^2_n}    \Lb   A^{\mbox{\tiny BFKL}}\Lb z\Rb\Rb^n
\\
\mbox{From \eq{LT6}}  &=&\sum^\infty_{n=0}\int^{2 \pi}_0\frac{ d \phi_1}{2 \pi}e^{ - i\,\phi_1,\,n} \, Z\Lb \phi_1 \Rb \int^{2 \pi}_0\frac{ d \phi_2}{2 \pi}e^{ - i\,\phi_2,\,n} \, Z\Lb \phi_2 \Rb \,n^2\,n! \Lb   -A^{\mbox{\tiny BFKL}}\Lb z\Rb\Rb^n \nn
\eea
Using formulae {\bf 8.357} and    {\bf 8.359} of Ref.\cite{RY} we can rewrite \eq{MPSI11} in the following form:
   \bea \label{MPSI12}
S\Lb z\Rb   &=& \frac{ d^2 }{d \ln \Lb A^{\mbox{\tiny BFKL}}\Lb z\Rb\Rb^2}  \int^{2 \pi}_0\frac{ d \phi_1}{2 \pi}\int^{2 \pi}_0\frac{ d \phi_2}{2 \pi}e^{ i ( \phi_1+\phi_2)} \, Z\Lb \phi_1 \Rb  \, Z\Lb \phi_2 \Rb\exp\Lb \frac{e^{ i ( \phi_1+\phi_2)} }{A^{\mbox{\tiny BFKL}}\Lb z\Rb}\Rb \,Ei\Lb - \frac{e^{ i ( \phi_1+\phi_2)} }{A^{\mbox{\tiny BFKL}}\Lb z\Rb}\Rb/A^{\mbox{\tiny BFKL}}\Lb z\Rb\nn\\
&=&\int^{2 \pi}_0\frac{ d \phi_1}{2 \pi}\int^{2 \pi}_0\frac{ d \phi_2}{2 \pi} \, Z\Lb \phi_1 \Rb  \, Z\Lb \phi_2 \Rb\Bigg\{\frac{e^{1/N} \left((N (N+3)+1) \Gamma \left(0,-\frac{1}{N}\right)-e^{1/N} N (2 N+3)\right)}{N^3}\,\,+\,\,{\cal O}\Lb \frac{1}{N}\Rb\Bigg\}
\eea   
with  $N  =  \exp\Lb - i ( \phi_1+\phi_2)\Rb A^{\mbox{\tiny BFKL}}\Lb z\Rb$. 

For large $ A^{\mbox{\tiny BFKL}}\Lb z\Rb$  \eq{MPSI12} reduces to the following expression:
  \beq \label{MPSI13} 
 S\Lb z\Rb\,\,=\,\, -\int^{2 \pi}_0\frac{ d \phi_1}{2 \pi}\int^{2 \pi}_0\frac{ d \phi_2}{2 \pi} \, Z\Lb \phi_1 \Rb  \, Z\Lb \phi_2 \Rb\frac{e^{-i \phi } \left(-\ln A^{\mbox{\tiny BFKL}}\Lb z\Rb+\ln \left(-e^{-i \phi }\right)+\gamma_E +2\right)}{A^{\mbox{\tiny BFKL}}\Lb z\Rb}  
 \eeq
  where $\phi = \phi_1 + \phi_2$ and $\gamma_E$ is   Euler's constant with numerical value  $\approx 0.577216$.   
  
   Using solution to \eq{LT12} (see \fig{sol}-b) we estimate the  behaviour  of S-matrix at large values of $z$:
    \beq \label{MPSI13} 
 S\Lb z\Rb\,\,= 0.987\frac{\ln\Lb A^{\mbox{\tiny BFKL}}\Lb z\Rb\Rb -\gamma_E -2}{A^{\mbox{\tiny BFKL}}\Lb z\Rb} \,\,+\,\,\frac{0.0274}{A^{\mbox{\tiny BFKL}}\Lb z\Rb} \eeq         
         
    \eq{MPSI13} is the main result of this paper which shows that the sum of large Pomeron loops  generates the scattering amplitude, which approaches its saturated value ($A\Lb z\Rb = 1$)  as $ exp\Lb- \h z\Rb$:
    \beq \label{MPSI14}
    A\Lb z\Rb\,\,=\,\,1\,\,-\,\,   0.987\frac{2}{N^2_0 \,z\,\exp\Lb \h z\Rb}
    \eeq

         ~
         
         ~
     \begin{boldmath}
     \section{Conclusions }
      \end{boldmath}

      
      In this paper we  summed the large BFKL Pomeron loops in the framework of CGC approach, using the leading twist approximation for the BFKL kernel. The main result of the paper is \eq{MPSI14}, which shows that the  sum of enhanced  BFKL Pomeron loop diagram generates the scattering amplitude which turns out to be much larger that in the case of deep inelastic scattering (see \eq{I1}). As we have discussed in the introduction  that this  result is  highly  unexpected and it contradicts (i)  the solution to the BK equation for the scattering amplitude that governs the deep inelastic scattering processes; (ii) 
    the results of Ref.\cite{IAMU}; and (iii)  the numerical simulations of Ref.\cite{MUSA}. However, the last contradiction has to be taken with a grain of salt since we have shown that the diffusion approximation  for the BFKL kernel which was used in Ref.\cite{MUSA} cannot provide the correct scattering amplitude at high energies\cite{LE1}. 
    
    As have been mentioned in the introduction we sincerely hope, that we made a mistake, which we failed to note,   and which our reader will  find. If not we need to reconsider our view on the sum of the BFKL Pomeron loops and accept that their summing will lead to large contribution of the rare    configurations in  CGC approach to the  scattering amplitude.

       {\bf Acknowledgements} 
     
   We thank our colleagues at Tel Aviv university  for
 discussions. Special thanks go A. Kovner and M. Lublinsky for stimulating and encouraging discussions on the subject of this paper. 
  This research was supported  by 
BSF grant 2022132.

\end{document}